\documentclass[reprint,NumberedRefs]{JASA}

\usepackage{graphicx}% Include figure files
\usepackage{dcolumn}% Align table columns on decimal point
\usepackage{bm}% bold math
\usepackage{amsmath,amssymb}
\usepackage{graphicx}
 \usepackage{color}
\usepackage[binary-units=true]{siunitx}
\usepackage{multirow}
\usepackage{braket}
\usepackage{longtable} 
\usepackage{setspace}

\newcommand{\ii}{\textrm{i}}
\newcommand{\ee}{\textrm{e}}

\newcommand{\dd}{\textrm{d}}
\newcommand{\bigO}{\textrm{O}}

\DeclareMathOperator{\re}{Re}
\DeclareMathOperator{\im}{Im}

\renewcommand{\vec}[1]{\bm{#1}}

\begin{document}

\title[Axisymmetric particle dynamics]
{Mean-acoustic fields  exerted on a subwavelength axisymmetric particle
}
\author{Everton B. Lima}
\author{Glauber T. Silva}\email{gtomaz@fis.ufal.br}
\affiliation{%
	Physical Acoustics Group,
	Instituto de F\'isica,
	Universidade Federal de Alagoas, 
	Macei\'o, AL 57072-970, Brazil}%

%\collaboration{MUSO Collaboration}%\noaffiliation

\date{\today}

\begin{abstract}
The acoustic radiation force produced by ultrasonic waves is the ``workhorse'' of particle manipulation in acoustofluidics.
Nonspherical particles are also subjected to a mean torque known as the acoustic radiation torque.
Together they constitute
\textit{the mean-acoustic fields} exerted on the particle.
Analytical methods alone cannot calculate these fields on arbitrarily shaped particles in actual fluids and no longer fit for purpose.
Here, a semi-analytical approach is introduced for handling subwavelength axisymmetric particles immersed in an isotropic Newtonian fluid.
The obtained mean-acoustic fields depend on
the scattering coefficients that reflect
the  monopole and dipole 
modes.
These coefficients are determined  by numerically solving the scattering problem.
Our method is benchmarked by comparison with the exact result for a subwavelength rigid sphere in water.
Besides, a more realistic case of a red blood cell immersed in blood plasma under a standing ultrasonic wave is investigated with our methodology.
\end{abstract}

%% pacs numbers not used

\maketitle

\section{Introduction}
Acoustofluidics is extensively based on the acoustic radiation force\cite{Torr1984} produced by ultrasonic waves to control and pattern micro/nanoparticles (cells, microorganisms, and viruses) in a liquid medium.
Nonspherical particles are also accompanied by the acoustic radiation torque.\cite{King1935}
The time-averaged radiation force and torque are referred to as \textit{the mean-acoustic fields} that drive the particle dynamics in most acoustofluidic settings.

The { analytical} solutions for the mean-acoustic fields depend, apart from the incoming wave, on the geometric shape\cite{Zhang2011a,Silva2014a,Nadal2016,Silva2018,Zhang2018,Gong2019,Gong2019a,Jerome2019,Lopes2020,Leao-Neto2020,Jerome2020,Gong2020,Leao-Neto2021} and material properties\cite{Hasegawa1969,
Silva2012,Mitri2012a,Baresch2013a,Leao-Neto2016,Leao-Neto2016a,Silva2019a} of the particle.
Moreover, the viscous\cite{Doinikov1994,Settnes2012,Zhang2014,Baasch2019} and thermodynamic\cite{Karlsen2015} properties of the surrounding fluid can significantly change the mean-acoustic fields on the particle.
The solutions mentioned above are found for particles with a simple geometric shape, such as spheres and spheroids.
{  Generally, these solutions are derived with the multipole expansion and angular spectrum methods.
The equivalence of these theoretical approaches have been demonstrated in Ref.~\onlinecite{Gong2021}. 
}

The mean-acoustic fields exerted on particles with more elaborate geometries are usually obtained through numerical techniques such as the finite element\cite{Glynne-Jones2013,Garbin2015,Schwarz2015} and boundary element\cite{Wijaya2015} method.
Nevertheless, numerical methods compute the mean-acoustic fields for a given  spatial configuration (position and orientation)
of the particle
relative to the incoming wave.
To determine  the particle position and orientation versus time,
the mean-acoustic fields have to be re-calculated at each time step in the interval of interest.
As a consequence, particle dynamics analysis in acoustofluidics might be a very intense computational task.
We emphasize that such analysis is of fundamental importance for understanding and designing micro/nanorobots propelled by ultrasound.\cite{Wang2012}

In acoustofluidics, subwavelength nonspherical particles, which are much smaller than the wavelength, have been conveniently modeled as small spheres mainly because
they fit the simple analytical expressions of the radiation force\cite{Gorkov1962} and torque.\cite{Silva2014a} obtained in the scattering dipole approximation.
Recently, analytical results considering a subwavelength spheroidal particle were also derived in the dipole approximation.\cite{Lima2020}
{ 
We stress that a complete theoretical description of the mean-acoustic fields for subwavelength particles of arbitrary geometry in an actual fluids seems to be unattainable. 
}

This article proposes a semi-analytical approach to describe the mean-acoustic fields interacting with an arbitrary axisymmetric particle in Newtonian thermoviscous fluids.
The obtained fields are related to the
pressure and fluid velocity of the incident wave and the
coefficients of the monopole and dipole modes of the scattered waves.
These coefficients are obtained from the projection of the scattered pressure onto the angular part of the corresponding mode (monopole and dipole). 
In turn, the scattered pressure is obtained via a finite element solver.
The numerical part of our method is verified against the exact solution for a rigid spherical particle in a lossless fluid.
Besides, we apply our method to compute the mean-acoustic fields exerted on a red blood cell by a standing plane wave in blood plasma.
The results, such as trap location and spatial orientation of the RBC, are in good agreement with previously reported experiments.

\section{Physical model}

\subsection{General assumptions}
Consider a subwavelength axisymmetric particle, i.e., with rotational symmetry around its principal axis, in a fluid of density $\rho_0$ and compressibility $\beta_0$.
The particle scatters an incoming wave of pressure amplitude $p_\text{in}$, angular frequency $\omega$,  wavelength $\lambda$, and wavenumber $k$.

\begin{figure}
    \centering
    \includegraphics[scale=.8]{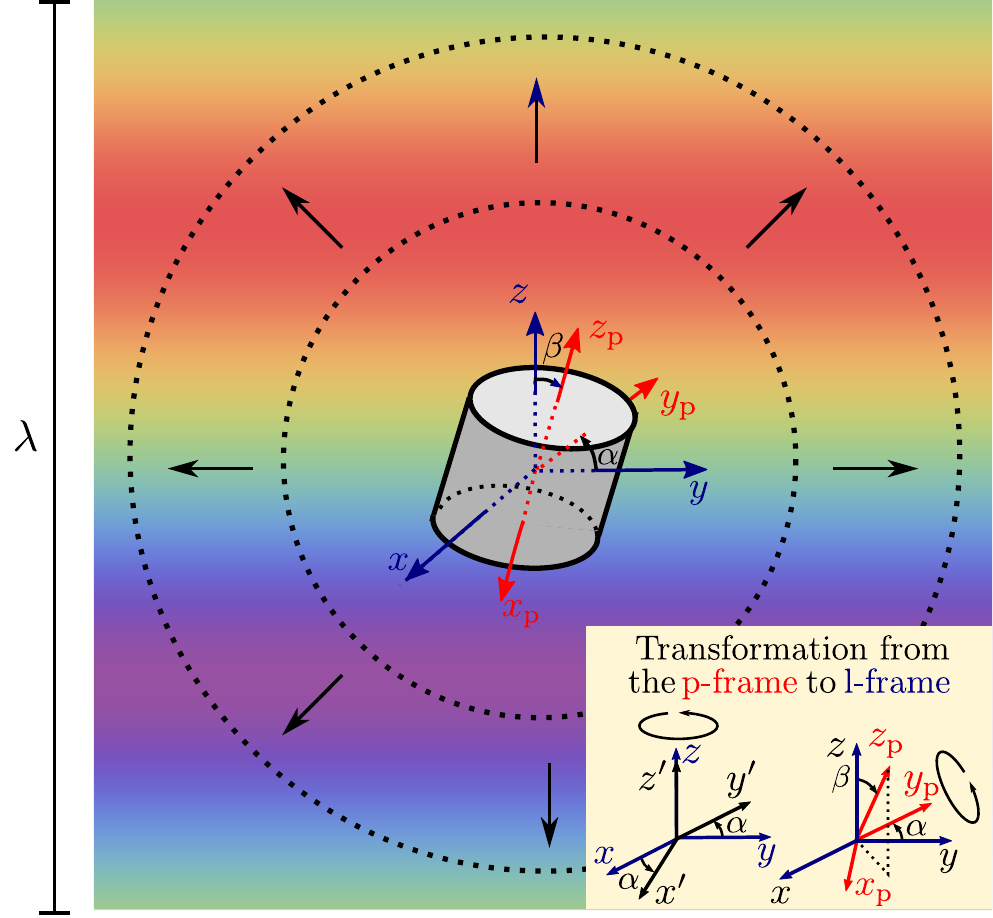}
    \caption{Problem sketch: an incoming wave (background image with a positive and negative crest in purple and red, respectively) of wavelength $\lambda$ is scattered by a small axisymmetric particle (gray cylinder). Two Euler angles $(\alpha, \beta)$ describe the particle orientation regarding a laboratory frame denoted by blue axes. 
    The particle frame is represented by
    red axes.
    The dotted concentric circles and black arrows denote the outgoing scattered wave.
    { 
    In the inset, we show the elementary rotations that map a vector from the particle frame (red axes) into the laboratory frame (blue axes).}
    }
    \label{fig:problem}
\end{figure}

{ 
In Fig.~\ref{fig:problem}, we sketch the
the acoustic scattering problem for an axisymmetric particle.
The scattering is conveniently described concerning the right-handed reference frame $O_{x_\text{p}y_\text{p}z_\text{p}}$ fixed in the geometric center of the particle (centroid), referred to as the p-frame. 
It is also useful to obtain the mean-acoustic fields in a fixed laboratory reference frame $O_{xyz}$ (l-frame) attached to the particle's centroid.
The cartesian unit vectors of both reference systems are
denoted by
$\vec{e}_i$ and $\vec{e}_{i_\text{p}}$, with
$i=x,y,z$ and $i_\text{p}=x_\text{p},y_\text{p},z_\text{p}$.
Since the particle is invariant under infinitesimal rotations
around the $z_\text{p}$ axis, 
we need two angles ($\alpha,\beta$), known as the Euler angles,  to transform a vector  from the p-frame to the l-frame.
The transformation structure is formed by two elementary rotations: a counterclockwise 
rotation of an angle $\alpha$ around the $z$ axis,  followed by a counterclockwise rotation of an angle $\beta$ around  new $y'$ axis--see Fig.~\ref{fig:problem}.
%
%
%In Fig.~\ref{fig:problem}, we note
%both p- and l-frame are right-handed co-ordinate systems.
%
We thus introduce the rotation tensor as
 \begin{align}
        \nonumber
        \mathbf{R} 
        &=
        \cos\alpha \cos\beta\, \vec{e}_x\vec{e}_{x_\text{p}}
        -\sin \alpha \, \vec{e}_x\vec{e}_{y_\text{p}}
        +
        \cos\alpha \sin\beta\, \vec{e}_x\vec{e}_{z_\text{p}}
        \\
        \nonumber
        &+\sin\alpha \cos\beta\, \vec{e}_y\vec{e}_{x_\text{p}} 
        +
        \cos \alpha \, \vec{e}_y\vec{e}_{y_\text{p}}
        +
        \sin\alpha \sin\beta\, \vec{e}_y\vec{e}_{z_\text{p}}
        \\
        &-\sin\beta\, \vec{e}_z\vec{e}_{x_\text{p}} 
        +
        \cos\beta\, \vec{e}_z\vec{e}_{z_\text{p}},
      \label{rotation}
    \end{align}
where
the product $\vec{e}_i\vec{e}_{i_\text{p}}$ is  a dyadic, which forms the standard basis of second rank tensors in the Euclidean space.
The Euler angles are given in the intervals
$0\le\alpha<2\pi$ and $0\le\beta\le \pi$.}
We define the particle orientation in both the p-frame and l-frame as 
\begin{subequations}
    \begin{align}
        \vec{e}_\text{p}   &=  \vec{e}_{z_\text{p}},\\
        \vec{e}_\text{l} &= {\bf R}\cdot \vec{e}_{z_\text{p}} = \cos \alpha\sin  \beta \, \vec{e}_x + \sin \alpha \sin \beta \, \vec{e}_y + \cos\beta\, \vec{e}_z,
        \label{Re}
    \end{align}
\end{subequations}
where the center dot represents the scalar product between a tensor and vector or two vectors.
{ 
Note also in deriving Eq.~\eqref{Re}, we have used the expressions 
$\vec{e}_i \vec{e}_{x_\text{p}}\cdot\vec{e}_{z_\text{p}} = \vec{e}_i \vec{e}_{y_\text{p}}\cdot\vec{e}_{z_\text{p}}=\vec{0}$,
and
$\vec{e}_i \vec{e}_{z_\text{p}}\cdot\vec{e}_{z_\text{p}}=\vec{e}_i$.
}
{ The inverse
transformation from  the l-frame to the p-frame is given in Appendix~\ref{app:rotation}.}

We assume that the largest dimension of the particle, denoted by $a$, is much smaller than the wavelength, $a\ll \lambda$, 
the so-called long-wavelength approximation.
To determine $a$ for an axisymmetric particle, we first notice that the particle occupies region, given in spherical coordinates, by
$\Omega_\text{p}= \{(r_\text{p},\theta_\text{p},\varphi_\text{p})\,|\,
0 \le r_\text{p}\le R_\text{p}(\theta_\text{p}), 0\le \theta_\text{p}\le \pi, 
0\le\varphi_\text{p}<2\pi\}$, with $R_\text{p}$ being the radial distance to the particle surface $S_\text{p}$.
The largest and smallest radial distance to the particle surface are, respectively,
\begin{subequations}
    \begin{align}
        a &= \max_{0\le\theta_\text{p}\le \pi}  R_\text{p}(\theta_\text{p}),\\
        b &= \min_{0\le\theta_\text{p}\le \pi}  R_\text{p}(\theta_\text{p}).
    \end{align}
\end{subequations}
In case of a spherical particle, $a=b$.
The characteristic size parameter of the subwavelength axisymmetric particle can be defined as
\begin{equation}
    k a=\frac{2\pi a}{\lambda}\ll 1.    
\end{equation}

\subsection{Scattering normal modes}

The scattered pressure by the particle is obtained by solving the Helmholtz equation with boundary conditions on the particle surface and the Sommerfeld radiation condition at the farfield ($\vec{r}_\text{p}\rightarrow \infty$).
The series solution represents the scattering pressure as an infinity sum of partial waves.
It is convenient to express the partial waves in spherical coordinates $(r_\text{p}, \theta_\text{p}, \varphi_\text{p})$.
Hence, the $n$th-order partial wave is
$h_n(k r_\text{p}) Y_n^m(\theta_\text{p},\varphi_\text{p})$, where 
$h_n$ is the spherical Hankel function of first type, 
$Y_n^m$ is the spherical harmonic, $m$ is the number representing the wave  orbital angular momentum.
In the case of subwavelength particles much smaller than the wavelength,
the partial wave expansion can be truncated at the dipole-moment term ($n=1$).
If we take a spherical particle as an example, we see that the truncation error is of the order of $\bigO[(ka)^2]$.
\cite{Leao-Neto2016a}  

In the dipole approximation, the partial wave expansion of the scattering pressure is
\begin{align}
    \nonumber
    p_\text{sc} &= p_0 \bigl[
    a_{00} s_{00}  Y_0^0(\theta_\text{p},\varphi_\text{p}) h_0(k r_\text{p})
    \\
    \nonumber
    &+ [a_{1,-1} s_{1,-1}Y_1^{-1}(\theta_\text{p},\varphi_\text{p})
    +a_{10} s_{10}
    Y_1^0(\theta_\text{p},\varphi_\text{p})\\
    &+a_{11} s_{11}
    Y_1^1(\theta_\text{p},\varphi_\text{p})]\,
    h_1(k r_\text{p}) 
    \bigr],
    \label{psc}
\end{align}
where $p_0$ and $a_{nm}$ are the pressure amplitude and beam-shape coefficient of the incoming wave, respectively.
We have four scattering coefficients representing  the monopole {  ($s_{00}$), transverse dipole ($s_{1,-1}$) and ($s_{11}$), and axial dipole ($s_{10}$) modes.}
In Eq.~\eqref{psc},
the time-dependent term $\ee^{-\ii \omega t}$  is omitted  for { simplicity}.

It is useful to introduce a function that represents the  angular distribution  of the $(n,m)$ mode on a spherical surface
of radius $R>a$ and centered at the particle centroid,
\begin{equation}
    \label{pnm}
    p_{nm}(\theta_\text{p},\varphi_\text{p}) = 
    \frac{p_\text{sc}(kR,\theta_\text{p},\varphi_\text{p})}{p_0 a_{nm} h_n(k R)}. 
\end{equation}
Using the orthogonality of the spherical harmonics, the scattering coefficients are given as the projection of the normalized scattered pressure onto the corresponding angular mode,
\begin{align}
    \nonumber
    s_{nm} &= \braket{n,m|p_{nm}}\\
    &=
    \int_0^{2\pi}\dd\varphi_\text{p} \int_0^\pi \dd \theta_\text{p}\, \sin \theta_\text{p} 
     \, p_{nm}(\theta_\text{p},\varphi_\text{p})
    Y_n^{m*}(\theta_\text{p},\varphi_\text{p}),
    \label{snm}
\end{align}
with asterisk denoting complex conjugation.
The Dirac's bra-ket notation $\braket{|}$ is used for simplicity.

In Appendix~\ref{app:symmetry}, we show that the transverse dipole modes of an
axisymmetric particle are degenerated, $s_{1,-1}=s_{11}$.
Thus,  the scattering coefficients of the problem are
    \begin{equation}
        \label{s-numerical}
        s_{00}  =\braket{0,0|{p}_{00}},\quad 
        s_{10}  =\braket{1,0|{p}_{10}}, \quad
        s_{11}  =\braket{1,1|{p}_{11}}.
    \end{equation}

\subsection{Acoustic radiation force}
The radiation force imparted on a subwavelength axisymmetric particle is given in terms of the scattering coefficients and incident fields as~\cite{Lima2020}
\begin{subequations}
    \label{Fradgeneral}
	\begin{align}
	\nonumber
    \vec{F}^\text{rad}_\text{p}
	& = \re\bigl[
	\bigl((\vec{e}_{x_\text{p}}\vec{e}_{x_\text{p}} 
	+ \vec{e}_{y_\text{p}}\vec{e}_{y_\text{p}})\mathcal{D}_{x_\text{p}y_\text{p}}  + \vec{e}_{z_\text{p}}\vec{e}_{z_\text{p}} \mathcal{D}_{z_\text{p}z_\text{p}}
	\bigr)
	\\
	\label{Frad}
	&\mbox{~~~~~~~~~}\cdot \vec{v}_\text{in,p}^*
	\bigr]_{\vec{r}_\text{p}=\vec{0}},\\
	\nonumber
	\mathcal{D}_{x_\text{p}y_\text{p}} &=-
	\frac{2\pi\ii }{k^2}
	\biggl[\frac{3\rho_0}{k}  \left(s_{11} 
	\vec{v}_\text{in,p}\cdot\nabla_{x_\text{p}y_\text{p}}
	+s_{10}v_{\text{in},z_\text{p}}\partial_{z_\text{p}}\right)\\
	&-\ii
	s_{00}\left(1+2s_{11}^*\right)
	\frac{p_\text{in}}{ c_0 }
	\biggr],\\
	\nonumber
	\mathcal{D}_{z_\text{p}z_\text{p}} & =-
	\frac{2\pi \ii }{k^2}\biggl[
	\frac{3\rho_0}{k}  \left(s_{11} \vec{v}_\text{in,p}\cdot
	\nabla_{x_\text{p}y_\text{p}}
	+s_{10}v_{\text{in},z_\text{p}}\partial_{z_\text{p}}\right)\\
	&
	-\ii
	s_{00}\left(1+2s_{10}^*\right)\frac{p_\text{in}}{ c_0 }
	\biggr].
	\end{align}
\end{subequations}
where
{ 
$\vec{v}_\text{in,p}=(v_{\text{in},x_\text{p}},v_{\text{in},y_\text{p}},v_{\text{in},z_\text{p}})$ is the fluid velocity regarding the p-frame,}
the asterisk denotes complex conjugation, and $\nabla_{x_\text{p}y_\text{p}} = \vec{e}_{x_\text{p}}\partial_{x_\text{p}}+\vec{e}_{y_\text{p}}\partial_{y_\text{p}}$.
The acoustic fields should be evaluated at  $\vec{r}_\text{p}=\vec{0}$.
We should bear in mind that the acoustic fields in \eqref{Fradgeneral} correspond to the incoming wave for which thermoviscous effects can be neglected.
Therefore, the lossless pressure-velocity relation 
\begin{equation}
    \label{pv}
    \vec{v} = - \frac{\ii \nabla p}{\rho_0 c_0 k}
\end{equation}
is used  into Eq.~\eqref{Frad} to obtain a simpler radiation force expression
\begin{align}
\nonumber
	&\vec{F}^\text{rad}_\text{p}
    = \\
	&\frac{1}{2} \re\biggl[\beta_0  \vec{\alpha}^{(\text{m})*}_\text{p}\cdot p_\text{in}^* \nabla_\text{p} p_\text{in}
	+ \rho_0\vec{\alpha}^{(\text{d})*}_\text{p} \cdot\vec{v}_\text{in,p}^* \cdot \nabla_\text{p} \vec{v}_\text{in,p} 
	\biggr]_{\vec{r}_\text{p}=\vec{0}}, 
	\label{Frad3}
\end{align}
where $\nabla_\text{p}=\vec{e}_{x_\text{p}}\partial_{x_\text{p}}+\vec{e}_{y_\text{p}}\partial_{y_\text{p}} + \vec{e}_{z_\text{p}}\partial_{z_\text{p}}$, and
\begin{subequations}
    \begin{align}
        \nonumber
        \vec{\alpha}_\text{p}^{(\text{m})} &= -\frac{4\pi \ii}{k^3}s_{00}\, [ {\bf I}
        +
        2 s_{11}^*(\vec{e}_{x_\text{p}}\vec{e}_{x_\text{p}}+\vec{e}_{y_\text{p}}\vec{e}_{y_\text{p}})\\
        &+
        2 s_{10}^*\vec{e}_{z_\text{p}}\vec{e}_{z_\text{p}}], 
        \\
        \vec{\alpha}_\text{p}^{(\text{d})} &=  -\frac{12\pi \ii}{k^3}\,[s_{11}(\vec{e}_{x_\text{p}}\vec{e}_{x_\text{p}} + \vec{e}_{y_\text{p}} \vec{e}_{y_\text{p}})+ s_{10}\vec{e}_{z_\text{p}}\vec{e}_{z_\text{p}}]
    \end{align}
\end{subequations}
are the monopole and dipole acoustic polarizability tensors (in units of volume) of the particle, and
 ${\bf I}=
\vec{e}_{x_\text{p}}\vec{e}_{x_\text{p}} +
\vec{e}_{y_\text{p}}\vec{e}_{y_\text{p}} + \vec{e}_{z_\text{p}}\vec{e}_{z_\text{p}}$
is the unit tensor.
Note that the radiation force
in Eq.~\eqref{Frad3} can also be understood in terms of the scattered and absorbed power by the particle in the dipole approximation (see Refs.~\onlinecite{Zhang2011,Mitri2014,Leao-Neto2017,Lopes2017}).

After deriving the radiation force in the p-frame, we can obtain the equivalent expression  in the l-frame. 
Using the rotational tensor described in Eq.~\eqref{rotation}, we find 
\begin{equation}
    \vec{F}^\text{rad} = \mathbf{R}\cdot \vec{F}^\text{rad}_\text{p}
\end{equation}
Using Eq.~\eqref{WPnabla} into this equation, we arrive at
\begin{subequations}
	\label{Frad4}
	\begin{align}
	\label{Fradfull}
	    \vec{F}^\text{rad}
	    &= \frac{1}{2} \re\biggl[\beta_0  \vec{\alpha}^{(\text{m})*}\cdot p_\text{in}^* \nabla p_\text{in}
	+ \rho_0\vec{\alpha}^{(\text{d})*} \cdot\vec{v}_\text{in}^* \cdot \nabla \vec{v}_\text{in} 
	\biggr]_{\vec{r}=\vec{0}}, \\
	\vec{\alpha}^{(\text{m})}&=
	{\bf R}\cdot\vec{\alpha}^{(\text{m})}_\text{p} \cdot {\bf R}^{-1},\\
	\vec{\alpha}^{(\text{d})}&={\bf R} \cdot \vec{\alpha}^{(\text{d})}_\text{p} \cdot{\bf R}^{-1},
	\end{align}
\end{subequations}
where $\vec{v}_\text{in}$ is the fluid velocity in the l-frame.
This exquisite expression describes the acoustic radiation force on a particle immersed in a Newtonian fluid concerning the l-frame.
Thermoviscous effects in the particle and fluid can be accounted for by solving the scattering problem considering temperature fields and appropriate boundary conditions.\cite{Karlsen2015}
Importantly, Eq.~\eqref{Frad4} is a generalization of the 
canonical form of the radiation force  derived for spherical particles.\cite{Settnes2012,Toftul2019}

We now proceed to the radiation force analysis of stationary and traveling waves.
As the pressure amplitude of a stationary  wave is a real-valued function, we have, $p_\text{in}^*\nabla p_\text{in} = (1/2)\nabla |p_\text{in}|^2$.
Also,  from Eq.~\eqref{pv} we see that 
$\vec{v}_\text{in}^* \cdot \nabla\vec{v}_\text{in}  = (1/2) \nabla |\vec{v}_\text{in}|^2$.
Thereby, for a stationary wave, Eq.~\eqref{Frad4} reduces to
a gradient force
\begin{equation}
    	\label{FradGrad}
    	\vec{F}^\text{rad} =
    	\left(\re\left[\vec{\alpha}^{(\text{m})}\right]\cdot
    	\nabla E_\text{pot} + \re\left[\vec{\alpha}^{(\text{d})}\right]
    	\cdot
    	\nabla E_\text{kin}\right)_{\vec{r}=\vec{0}}
\end{equation}
This equation shows that the kinetic $E_\text{kin}=\rho_0 |\vec{v}_\text{in}|^2/4$ and potential $E_\text{pot}=\beta_0|p_\text{in}|^2/4$ energy densities pumped into the fluid are transformed into the acoustic force of radiation through their gradient projection onto the real part of the particle polarizabilities.

A nearly ubiquitous stationary wave employed in acoustofluidic devices is a standing plane wave whose acoustic fields are
\begin{subequations}
\label{swfields}
    \begin{align}
    \label{p_standing}
        p_\text{in} &= p_0 \cos [k (z+z_0)],\\
        \vec{v}_\text{in} &=
         \frac{\ii p_0}{\rho_0 c_0} \sin [k (z+z_0)]\, \vec{e}_z,
         \label{vinz}
    \end{align}
\end{subequations}
where $z_0$ is the particle position relative to the pressure antinode at $z=0$.
Replacing ~\eqref{swfields} into Eq.~\eqref{FradGrad}, we obtain the radiation force as
\begin{subequations}
\label{FradSW}
    \begin{align}
        \vec{F}^\text{rad} &= F_0
        \mathbf{\Phi} \sin 2 k z_0,\\
        F_0 &= A E_0, \quad
        E_0 = \frac{1}{4}\beta_ 0 p_0^2,
    \end{align}
\end{subequations}
where $E_0$ is  the acoustic energy density, and
$F_0$ is the characteristic force.
The characteristic area $A$ can be either the axial ($A_{||}$) and transverse ($A_\perp$) particle cross-section area.
The acoustophoretic vector is expressed by
\begin{align}
    \nonumber
    &\mathbf{\Phi} = \frac{k}{ A}\re\left[\vec{\alpha}^{(\text{m})}
    - \vec{\alpha}^{(\text{d})}\right]\cdot \vec{e}_z  \\
    \nonumber
    &
    =\frac{2\pi}{k^2A}\cos \alpha \sin 2\beta\, 
    \re[3\ii( s_{10} -  s_{11}) -2\ii s_{00} ( s_{10}^* - s_{11}^*)]
    \vec{e}_x\\
    \nonumber
    &+ \frac{2\pi}{k^2A}\sin \alpha \sin 2\beta\, 
    \re[3\ii (s_{10} -  s_{11}) -2\ii(  s_{10}^* -  s_{11}^*)]
    \vec{e}_y\\
    \nonumber
    &+\frac{4\pi}{k^2A} \re[3\ii(s_{10}\cos^2\beta + s_{11} \sin^2 \beta)
    - \ii s_{00} (1 + 2 s_{10}^* \cos^2\beta \\
    &+ 2 s_{11}^* \sin^2\beta)]
    \vec{e}_z.
    \label{Phi}
\end{align}
We see the radiation force depends on the particle orientation angles $\alpha$ and $\beta$.
In this regard, we shall see later that the possible equilibrium configurations are the axial ($\beta=0$) and transverse orientation ($\beta=\pi/2$).
In either orientation, the radiation force becomes an axial force,
\begin{equation}
    \label{FSW}
    \vec{F}^\text{rad} =F_0  \Phi \sin 2 k z_0\, \vec{e}_z,
\end{equation}
where $\Phi$ jointly represents the axial
$\Phi_{||}$ and transverse $\Phi_\perp$ acoustophoretic factors.
They  are  expressed by 
\begin{subequations}
\label{phi_SW}
    \begin{align}
        \label{phi_axial_SW}
        \Phi_{||} &= \frac{4 \pi}{k^2A_{||}}\re[3\ii s_{10} - \ii s_{00}(1+ 2 s_{10}^*)], \quad (\beta = 0),\\
        \Phi_\perp &=\frac{4\pi}{k^2A_\perp} \re[3\ii s_{11} - \ii s_{00}(1+ 2 s_{11}^*)], \quad (\beta = \pi/2).
        \label{phi_perp_SW}
    \end{align}
\end{subequations}
The position where the particle will be trapped depends on the sign of  $\Phi$.
The trap is located in a pressure node if $\Phi>0$, and in a pressure antinode, otherwise. 
For a spherical particle, the axial and transverse dipole modes are degenerate $s_{10}=s_{11}$--see Eq.~\eqref{s-sphere}. 
Hence, the axial and transverse acoustophoretic factors equalize,
$\Phi_{||}=\Phi_\perp$.
This result is also  in agreement with previously obtained solutions.\cite{Gorkov1962,Bruus2012,Silva2014a} 
The equations in \eqref{phi_SW} also agree with the analytical expressions derived for a prolate spheroidal particle.\cite{Silva2018}

Now we focus our analysis on a traveling plane wave with pressure and velocity fields given by
\begin{subequations}
    \begin{align}
        p_\text{in}&= p_0 \ee^{\ii k z},\\
        \vec{v}_\text{in} &=  \frac{p_0}{\rho_0 c_0} \ee^{\ii k z}\, \vec{e}_z.
        \label{vpw}
    \end{align}
\end{subequations}
In this particular case, the radiation force in Eq.~\eqref{Fradfull} becomes
	\begin{align}
	    \nonumber
	    &\vec{F}^\text{rad}
	    =F_0 \vec{\Phi},\\
	&\vec{\Phi} = -\frac{2k}{ A} \im\left[\left(\vec{\alpha}^{(\text{m})} + \vec{\alpha}^{(\text{d})}\right)  
	\right]\cdot \vec{e}_z\\
	\nonumber
    &
    =\frac{4\pi}{k^2 A}\cos \alpha \sin 2\beta\, 
    \re[3 (s_{10} -  s_{11}) +2s_{00} ( s_{10}^* -   s_{11}^*)]
    \vec{e}_x\\
    \nonumber
    &+ \frac{4\pi}{k^2 A}\sin \alpha \sin 2\beta\, 
    \re[3 (s_{10} -  s_{11}) +2s_{00} ( s_{10}^* -   s_{11}^*)]
    \vec{e}_y\\
    \nonumber
    &+8\pi\re[3(s_{10}\cos^2\beta + s_{11} \sin^2 \beta)
    + s_{00} (1 + 2 s_{10}^* \cos^2\beta \\
    &+ 2 s_{11}^* \sin^2\beta)]
    \vec{e}_z.
\end{align}
Here the acoustophoretic vector is a result of the scattered and absorbed momentum fluxes.\cite{Leao-Neto2017,Lopes2017}
The radiation force dependence with particle orientation is also noted.
Again, the possible axial and transverse equilibrium orientations of the particle are $\beta=0,\pi/2$, which leads to an axial radiation force, 
\begin{equation}
    \vec{F}^\text{rad}
	    =F_0
	    {\Phi}\vec{e}_z.
\end{equation}
The  factor $\Phi$ jointly represents
\begin{subequations}
    \label{PhiPW}
    \begin{align}
        \label{phi_axial_TW}
        \Phi_{||}  &= 8\pi\re[ 3 s_{10} + s_{00} (1 + 2s_{10}^*)],  \quad (\beta = 0),\\
        \Phi_\perp &= 8\pi\re[ 3 s_{11} + s_{00} (1 + 2s_{11}^*)], \quad (\beta = \pi/2).
        \label{phi_perp_TW}
    \end{align}
\end{subequations}
It is worth noticing that \eqref{PhiPW} covers previous results for a spherical particle.\cite{Gorkov1962,Bruus2012, Silva2014a}

\subsection{Acoustic radiation torque}
\label{Sec:torque}
The acoustic radiation torque $\vec{\tau}^\text{rad}_\text{p}$ generated by an acoustic wave 
relative to the p-frame
is expressed by\cite{Lopes2020,Lima2020}
\begin{subequations}
	\begin{align}
	\nonumber
	\vec{\tau}^\text{rad}_\text{p}
		& =  \frac{6 \pi \rho_0 }{k^3}
	\text{Im} 
	\biggl[
	\gamma_{\perp}
	\left(
	v_{\text{in},y_\text{p}}  \vec{e}_{x_\text{p}} 
	-v_{\text{in},x_\text{p}} \vec{e}_{y_\text{p}}
	\right)v_{\text{in},z_\text{p}}^*
	\\
	&+\gamma_{||} v_{\text{in},x_\text{p}} v_{\text{in},y_\text{p}}^*\,\vec{e}_{z_\text{p}}\biggr]_{\vec{r}_\text{p}=\vec{0}},
    \label{radtorque_p}\\
    \label{gamma_perp}
    \gamma_{\perp} &= s^*_{10}+s_{11}+2s_{11}s^*_{10},\\
    \gamma_{||} &= s_{11}+s_{11}^*+2|s_{11}|^2.
	\end{align}
\end{subequations}
The quantities $\gamma_\perp$ and $\gamma_{||}$ are
the transverse and axial gyroacoustic functions.
To obtain the radiation torque in the l-frame, we first need to consider the velocity field relation
\begin{equation}
    \label{vp}
    \vec{v}_\text{in,p} = {\bf R}^{-1}\cdot \vec{v}_{\text{in}}.
\end{equation}
Using \textsc{Mathematica} software (Wolfram, Inc., USA), we  insert Eq.~\eqref{vp} into Eq.~\eqref{radtorque_p}
and apply the rotational tensor to obtain
\begin{widetext}
\begin{align}
        \nonumber
 \vec{\tau}^\text{rad} &= {\bf R}\cdot \vec{\tau}_\text{p}^\text{rad}\\
        \nonumber
       & =\frac{6 \pi \rho_0 }{k^3}
	\text{Im} 
	\biggl\{\bigl[
	 \gamma_\perp (v_{\text{in},y} \cos \beta - v_{\text{in},z} \sin \alpha \sin \beta )
	 \left(v_{\text{in},x}^* \cos \alpha  \sin \beta + v_{\text{in},y}^* \sin \alpha  \sin \beta + v_{\text{in},z}^* \cos \beta \right) 
	 \\
    \nonumber
     &+ \gamma_{||} \cos \alpha  \sin \beta  \left(v_{\text{in},y}^* \cos \alpha - v_{\text{in},x}^* \sin \alpha \right) (v_{\text{in},x} \cos \alpha  \cos \beta + v_{\text{in},y} \sin \alpha \cos \beta - v_{\text{in},z} \sin \beta )\bigr]\,\vec{e}_x
	 \\
	 \nonumber
	 &+\bigl[
	\gamma_{||} \sin \alpha  \sin \beta  \left(v_{\text{in},y}^* \cos \alpha -v_{\text{in},x}^* \sin \alpha \right) (v_{\text{in},x} \cos \alpha  \cos \beta+ v_{\text{in},y} \sin \alpha \cos \beta - v_{\text{in},z} \sin \beta )\\
	\nonumber
	&-\gamma_\perp (v_{\text{in},x} \cos \beta 
	-v_{\text{in},z} \cos \alpha \sin \beta) \left( v_{\text{in},x}^* \cos \alpha \sin \beta + v_{\text{in},y}^* \sin \alpha \sin \beta + v_{\text{in},z}^* \cos \beta \right)
	 \bigr]\,\vec{e}_y\\
	 \nonumber
	 &+ \bigl[
        \gamma_\perp \sin \beta (v_{\text{in},x} \sin \alpha - v_{\text{in},y} \cos \alpha ) \left(v_{\text{in},x}^* \cos \alpha \sin \beta + v_{\text{in},y}^* \sin \alpha  \sin \beta  +  v_{\text{in},z}^* \cos \beta \right)
        \\
        &+ \gamma_{||} \cos \beta \left(v_{\text{in},y}^* \cos \alpha - v_{\text{in},x}^* \sin \alpha \right) 
        (v_{\text{in},x} \cos \alpha  \cos \beta + v_{\text{in},y} \sin \alpha  \cos \beta - v_{\text{in},z} \sin \beta ) \bigr]\,\vec{e}_z
	\biggr\}_{\vec{r}=\vec{0}}.
        \label{radtorque_wave}
\end{align}
\end{widetext}

We move on to analyze the acoustic radiation torque produced by a standing plane wave
along the $z$ axis as described in~\eqref{swfields}.
With no transverse fluid velocity components, $v_{\text{in},x}=v_{\text{in},y}=0$, the radiation torque in \eqref{radtorque_wave} becomes
\begin{subequations}
    \begin{align}
        \label{tau_z}
        \vec{\tau}^\text{rad} &= \frac{3\pi \rho_0 }{k^3} 
        \im[\gamma_\perp] |v_{\text{in},z}|^2\sin 2\beta\,
        \vec{e}_\alpha,\\
        \vec{e}_\alpha &= \cos\alpha\,\vec{e}_y -   \sin \alpha\,\vec{e}_x.
    \end{align}
\end{subequations}
The unit vector $\vec{e}_\alpha$ is orthogonal to the particle axis of symmetry, $\vec{e}_\alpha \cdot \vec{e}_\text{l} = 0$. 
Substituting Eq.~\eqref{vinz} into Eq.~\eqref{tau_z} yields
\begin{subequations}
    \label{RTSW2}
    \begin{align}
    \vec{\tau}^\text{rad} 
    &= \tau_0 \gamma^\text{i}_\perp  \sin^2 k z_0 \sin 2\beta \, \vec{e}_\alpha,
    \label{RTSW}\\
    \tau_0 &= V E_0, \quad \gamma^\text{i}_\perp = 
    \frac{12 \pi  }{k^3 V} \im[\gamma_\perp],
\end{align}
\end{subequations}
where $\gamma^\text{i}_\perp$ is the transverse gyroacoustic parameter.
In acoustofluidics, the typical characteristic torque scale is $\tau_0 \sim  \SI{1}{\nano\newton \micro\meter}$.

As the particle will be trapped in either a pressure node or antinode, the pre-sine factor is reduced to one, $\sin^2 k z_0=1$.
So  the particle will be transversely aligned ($\beta=\pi/2$) to the wave axis as
the  gyroacoustic parameter is positive, $\gamma^\text{i}_\perp>0$.
The axial orientation parallel to the wave axis ($\beta=0$) occurs as $\gamma^\text{i}_\perp<0$.

Let us switch gears toward the radiation torque due to a traveling plane wave.
By replacing Eq.~\eqref{vpw} into the radiation torque expression in \eqref{radtorque_wave}, we find
\begin{equation}
    \vec{\tau}^\text{rad} =\tau_0 \gamma^\text{i}_\perp  \sin 2\beta 
    \,\vec{e}_\alpha.
    \label{RTPW}
\end{equation}
Apart from the pre-sine factor $\sin^2 k z_0$, this result is identical to the radiation torque by a standing wave given in Eq.~\eqref{RTSW}.
Therefore, the particle orientation follows the same conclusion as the standing wave case.

\section{Finite element model}
To compute the scattering coefficients given in \eqref{s-numerical}, we numerically solve the scattering of a traveling plane wave using the finite element (FE) method in the commercial software \textsc{Comsol} Multiphysics  ({Comsol}, Inc., USA). 
The choice for plane wave scattering is justified
due to its simplicity and for having
analytical expressions of the corresponding
beam-shape coefficients.
{ 
Once the scattering coefficients are determined, they can be used in conjunction with any incident wave.
}
The p-frame is adopted in the finite element model.
Besides, we use the \textit{Acoustics Module} and \textit{Solid Mechanics Module}, where the equations of linear acoustics and elastodynamics are implemented.

The pressure amplitude $p_\text{in}=p_0 \ee^{\ii \vec{k}\cdot \vec{r}_\text{p}}$, with wavevector $\vec{k} = k(\sin\beta \,\vec{e}_{x_\text{p}} + \cos\beta\,\vec{e}_{z_\text{p}}),$ is set as the background pressure in the computational domain.
To obtain the scattering coefficients in \eqref{s-numerical}, we need the monopole and dipole beam-shape coefficients of a traveling plane wave,\cite{Colton1998}
\begin{equation}
    a_{00} = 2\sqrt{\pi}, \quad
    a_{10} = 2 \ii \sqrt{3\pi} \cos \beta, \quad
    a_{11} = -\ii \sqrt{6 \pi} \sin \beta.   
\end{equation}

For a rigid particle,  the  normal component of the fluid velocity should vanish at the particle surface $S_0$,
\begin{equation}
(\vec{v}_\text{in,p}+\vec{v}_\text{sc,p})|_{{\bm r}_\text{p}\in S_0} \cdot \vec{n} = 0, 
\end{equation} 
where $\vec{v}_\text{sc,p}$ is the scattered fluid velocity and  $\vec{n}$ is the outward normal vector of the particle.
In the case of an elastic solid particle,  the boundary conditions across the particle surface are 
the continuity of normal stresses and displacements, and zero tangential stress condition.
Accordingly, we have
\begin{subequations}
    \begin{align}
        ({p}_\text{in}+{p}_\text{sc} - \vec{n}\cdot \vec{\sigma}_\text{p} )|_{{\bm r}_\text{p}\in S_0}&=0,\\
        \vec{n} \cdot \left[\frac{\ii}{\omega}(\vec{v}_\text{in,p}+\vec{v}_\text{sc,p})- \vec{u}_\text{p}\right]_{{\bm r}_\text{p}\in S_0}  &=0,\\
        \vec{t} \cdot \vec{\sigma}_\text{p}|_{{\bm r}_\text{p}\in S_0} &= 0.
    \end{align}
\end{subequations}
where $\vec{t}$ is the tangential unit vector,
$\vec{\sigma}_\text{p}$ and $\vec{u}_\text{p}$ are the stress tensor and displacement vector inside the particle, respectively. 
\LTcapwidth=.48\textwidth
\begin{longtable}{lr}
\caption{
\label{tab:parameters}
Geometrical and physical parameters used in the finite element simulations of the scattering problem.
}\\
\hline
\multicolumn{1}{l}{\bf Parameter}   &  
\multicolumn{1}{c}{\bf Description} \\
\hline\hline
Computational domain & Cylindrical \\
    Diameter and height & $100 a$\\
    Radius of the integration surface ($R$) & $5a/4$ \\
    Minimum element size ($e_\text{min}$) & $ a/200$ \\
    Maximum element size ($e_\text{max}$) & $a/10$ to $a/3$ \\
    Mesh growth rate & $10\,\%$\\
    PML type & Polynomial\\
    PML scaling factor & $1$\\
    PML curvature parameter & $2$\\
    PML thickness & $50 a$ \\
    PML layers & $30$--$50$ \\
     Frequency & $2$--$\SI{12}{\mega\hertz}$\\
    \hline
    \hline
    Fluid medium  & {Water}\\
    Mass density ($\rho_0$) & $\SI{998}{\kilogram\per\meter\cubed}$ \\
    Compressibility ($\beta_0$) & $\SI{0.4560}{\per\giga\pascal}$   \\
    \hline
    Particle &  Rigid sphere\\
    Radius ($a$) & $\SI{3.91}{\micro\meter}$\\
\hline 
\hline
    Fluid medium  & {Blood plasma}\cite{Gupta2018}\\
    Mass density ($\rho_0$)         & $\SI{1026}{\kilogram\per\meter\cubed}$          \\
    Compressibility ($\beta_0$)      & $\SI{0.4077}{\per\giga\pascal}$         \\
\hline
    Particle &  Red blood cell\\
    Mass density\cite{Gupta2018} ($\rho_\text{p}$)  & $\SI{1100}{\kilogram\per\meter\cubed}$   \\
    Young's modulus\cite{Dulinska2006} ($E$) & $\SI{26}{\kilo\pascal}$\\
    Compressibility\cite{Gupta2018} ($\beta_\text{p}$) & $\SI{0.34}{\per\giga\pascal}$\\
    Geometric parameters:\cite{Evans1972} &\\
    Major semiaxis ($a$) & $R_0= \SI{3.91}{\micro\meter}$ \\ 
    Minor semiaxis ($b$) & $C_0/2=\SI{0.405}{\micro\meter}$ \\ 
    Constant $C_2$ & $\SI{7.83}{\micro\meter}$ \\  
    Constant $C_4$ &$-\SI{4.39}{\micro\meter}$\\
\hline
\end{longtable}

The  fluid domain  corresponds to a mesh within a cylindrical region. 
We use a perfectly matched layer (PML) in a concentric cylindrical shell of a quarter-wavelength thickness to absorb the outgoing scattered wave,
and thus avoid reflection back into the fluid domain.
At the fluid-PML interface, the acoustic fields are continuous.
The outer surface of the PML is a rigid wall.

To proceed with the FE modeling, we need
to choose the appropriate mesh symmetry:
2D axisymmetric or 3D model.
The 2D axisymmetric model is less computationally intense and corresponds to a plane wave propagating along the particle symmetry axis with $\beta=0$.
It can be is used to compute the
monopole ($s_{00}$) and axial ($s_{10}$) dipole coefficients only.
The 2D mesh model is not suitable for computing the transverse dipole coefficient ($s_{11}$), because $\beta=0$ implies  $a_{11} = 0$, which leads to a not well-defined outcome of Eq.~\eqref{pnm}.
We are, in principle, left with a full 3D scattering model to find $s_{11}$, which usually do not fit in low-memory computers ($<\SI{16}{\giga\byte}$).
Nevertheless, the \textsc{Comsol} FE solver transforms a 3D acoustic scattering by an axisymmetric object into a collection of 2D scattering problems using the built-in \textit{Plane Wave Expansion} method.\cite{Comsol2021}
We, therefore, use this method with five terms in the plane wave expansion to compute $s_{11}$ with $\beta=\pi/2$.
Here, the acoustic scattering by a rigid spherical particle and a red blood cell (RBC) is solved via the FE model.
The simulation parameters used in the numerical solutions are presented in Table~\ref{tab:parameters}.

In the upcoming analysis, the RBC is modeled as a solid elastic material with 
a biconcave disk-shaped geometry that is described in cylindrical coordinates $(\varrho,\varphi,z)$ by the implicit equation\cite{Evans1972} 
\begin{equation}
 z^2 = \frac{1}{4}\left[1 - \left(\frac{\varrho}{R_0}\right)^2\right]\left[C_0 + C_2\left(\frac{\varrho}{R_0}\right)^2 + C_4\left(\frac{\varrho}{R_0}\right)^4\right]^2,
\end{equation}
where $2R_0$ and $C_0$ are, respectively, the cell transverse diameter and central thickness, and $C_2$, and $C_4$ are geometric parameters given in Table~\ref{tab:parameters}. 
\begin{figure}
    \centering
    \includegraphics[scale=1.67]{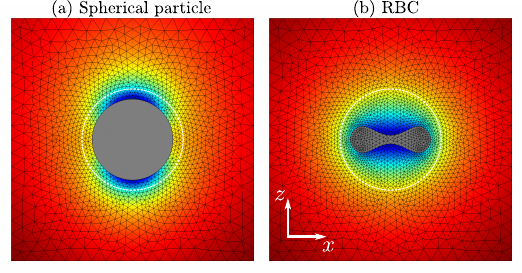}
    \caption{\label{fig:mesh}
    The 2D axisymmetric mesh for the plane wave scattering by (a) a rigid spherical particle in water and (b) a RBC (modeled as an elastic solid) in blood plasma.
    The coarse mesh is for illustration purposes only.
    The particles, shown in gray, have the same diameter $2a=\SI{7.82}{\micro\meter}$.
    The white dashed circle illustrates the integration surface; upon it, the scattering coefficients are computed.
    The plane wave travels in the vertical direction along the $z$ axis. 
    The background image corresponds to the real part of the computed scattered wave, where blue and red regions mean negative and positive pressures, respectively.
    }
\end{figure}

Now we define the  mesh convergence parameter as
\begin{align}
    \nonumber
    \epsilon &\equiv\\
    &\max\left\{\frac{\left|\re \left[s_{nm}^\text{ref}\right] - \re\left[s_{nm}\right] \right|}{\left|\re \left[s_{nm}^\text{ref}\right]\right|}, \frac{\left|\im \left[s_{nm}^\text{ref}\right] - \im\left[s_{nm}\right] \right|}{\left|\im \left[s_{nm}^\text{ref}\right]\right|}\right\}.
    \label{error}
    \end{align}
{ The reference solution $s_{nm}^\text{ref}$ 
for the spherical particle is related to the exact coefficients given by} 
\begin{equation}
    \label{s-sphere}
        s_{00} =  -\frac{j_0'(ka)}{h_0'(ka)},\quad
        s_{10} = s_{11} = -\frac{j_1'(ka)}{h_1'(ka)},
\end{equation}
where $j_n$ is the spherical Bessel function of order $n$, { and the prime symbol means differentiation.}
Whereas for the RBC, $s_{nm}^\text{ref}$ is the solution
that shows negligible changes after further mesh refinements.
The FE model was executed on a \textsc{Supermicro} workstation ({Supermicro} Computer, Inc., USA) based in a dual-processor Intel Xeon E5-2690v2 @ $\SI{3}{\giga\hertz}$ and $\SI{224}{\giga\byte}$  memory size.
The computational time is about $\SI{1}{\minute}$ in the 2D mesh and
$\SI{5}{\minute}$ with the plane wave expansion method.
\begin{figure}
    \centering
    \includegraphics[scale=.32]{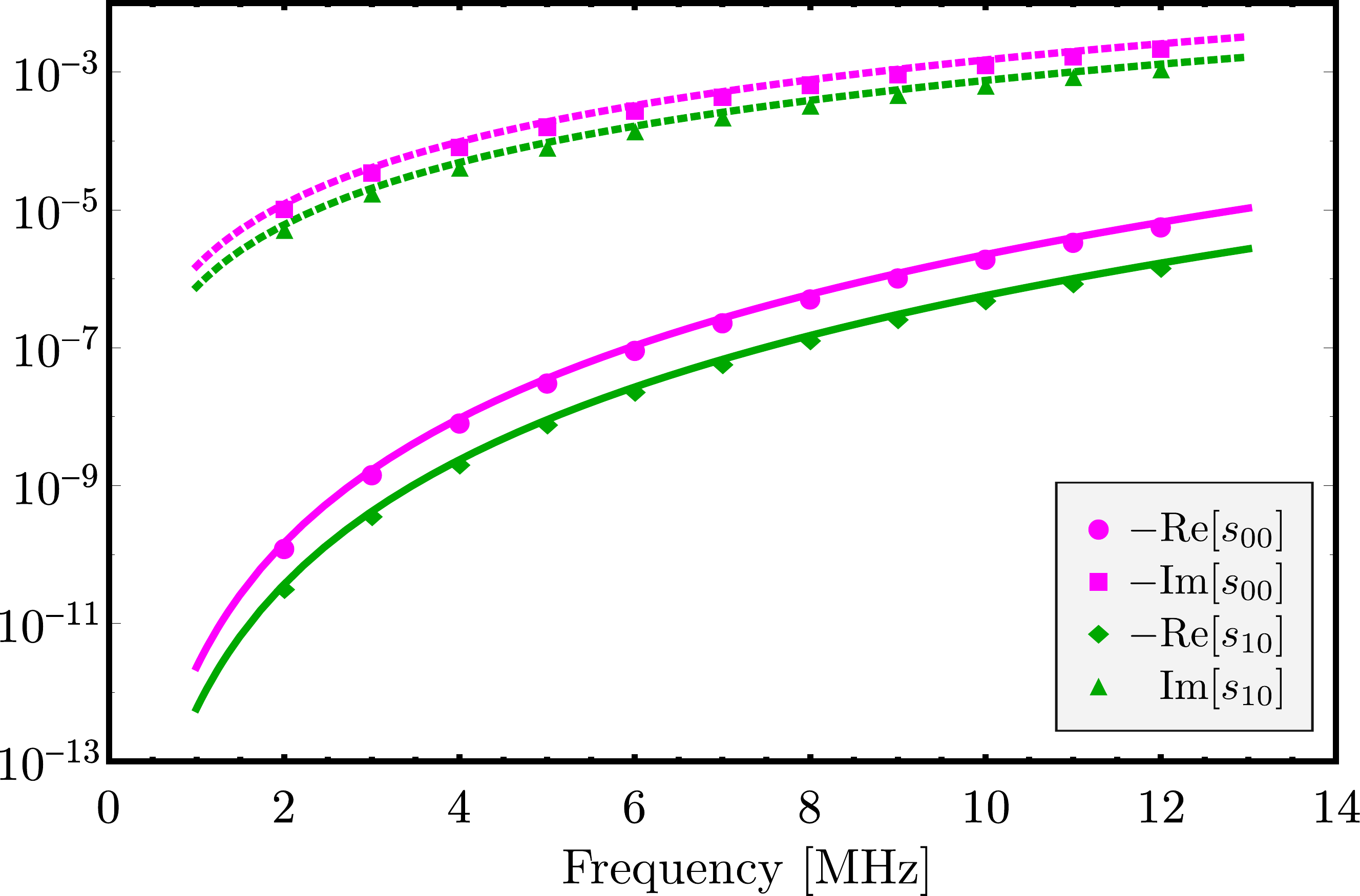}
    \caption{\label{fig:sphere}
    The scattering coefficients of a rigid spherical particle much smaller than the wavelength.
    The solid and dashed lines are
    the corresponding exact solution from~\eqref{s-sphere}.
    }
\end{figure}

\section{Results and discussion}

In Fig.~\ref{fig:mesh}, we depict the 2D axisymmetric mesh domain around the spherical particle and RBC.
The particles have same diameter  $2a=\SI{7.82}{\micro\meter}$. 
The integration surface, where the scattering coefficients are computed, is shown as a white circle enclosing the particle.
The real part of the scattered pressure is also illustrated
as the background image.
The positive and negative pressure correspond to blue and yellow-to-red regions, respectively.

In Fig~\ref{fig:sphere}, we plot scattering coefficients of the small rigid sphere in water for a routinely  frequency range in acoustofluidics,  
$2$--$\SI{12}{\mega\hertz}$.
The numerical result is in excellent agreement with the exact solution.
At $\SI{2}{\mega\hertz}$,
the convergence parameter is $\epsilon=7\,\%$,
with the maximum element size  $e_\text{max}=a/3$. 
This value drops to $\epsilon=1\,\%$ at $\SI{3}{\mega\hertz}$, and keeps going down as the frequency increases.
In our numerical studies, the solution convergence becomes an issue at frequencies smaller than $\SI{1}{\mega\hertz}$.
We also found that the transverse and axial dipole coefficients are degenerated, $s_{11}=s_{10}$, within the numerical error.
{ 
We note that the convergence error as calculated from Eq.~\eqref{error} is more prominent in the real part of the scattering coefficient, for which the reference solution is of the order  $s_{nm}^\text{ref}=\bigO[(ka)^6]$.\cite{Leao-Neto2016a} 
So, as the frequency decreases, the reference solution is drastically reduced, which may cause a convergence instability at lower frequencies.
}
\begin{figure}
    \centering
    \includegraphics[scale=.45]{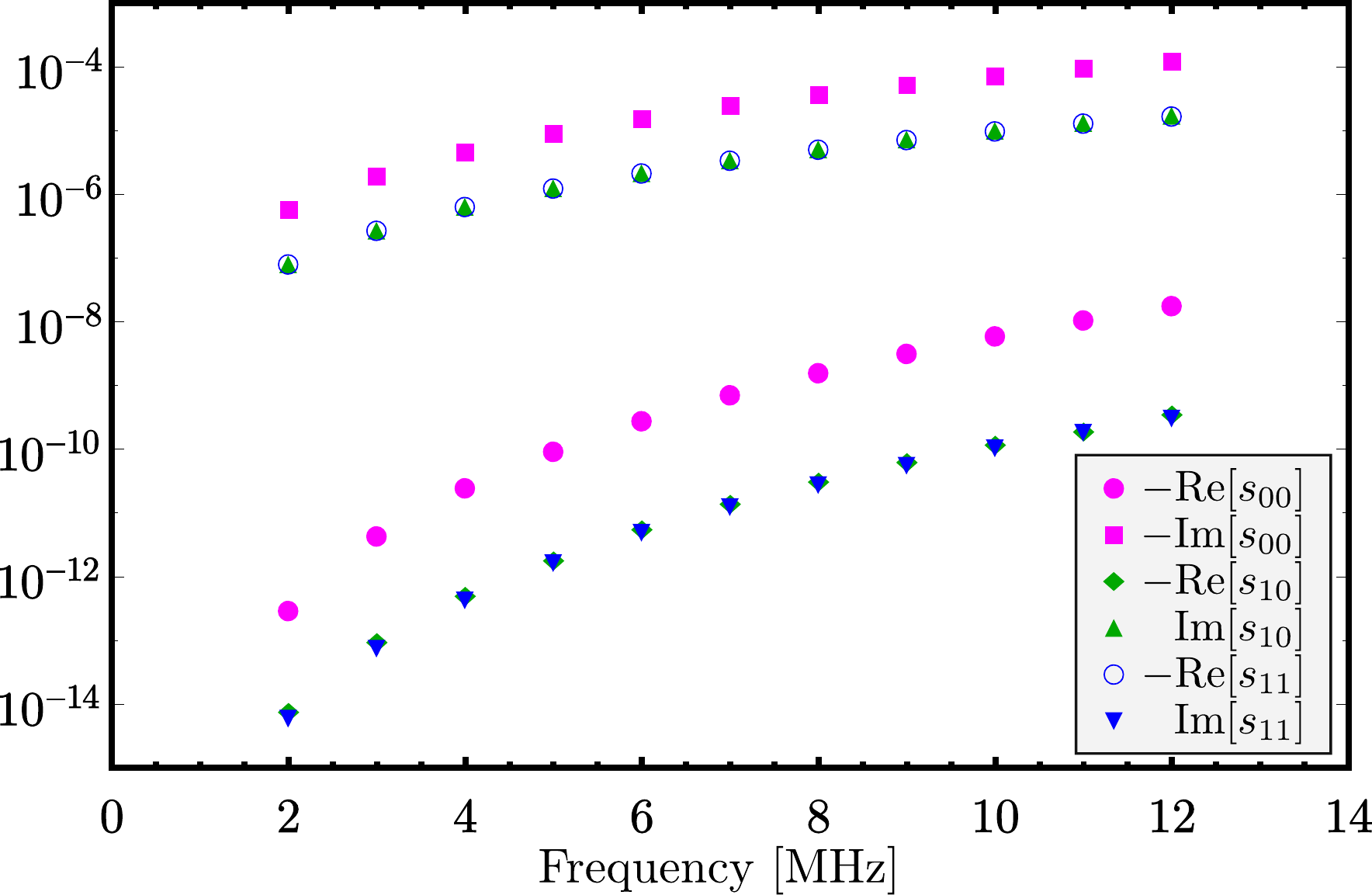}
    \caption{\label{fig:RBC}
    The scattering coefficients of a RBC in blood plasma  versus frequency.
    }
\end{figure}

 Figure~\ref{fig:RBC} shows the scattering coefficients of the RBC in blood plasma.
In this study, the mesh convergence parameter is $\epsilon<9\,\%$, for the maximum element size  $e_\text{max}=a/10$. 
The parameter approaches the upper bound at lower frequencies.
With these coefficients, we can proceed to compute the radiation force and torque generated by a standing plane wave.

\begin{figure}
    \centering
    \includegraphics[scale=.47]{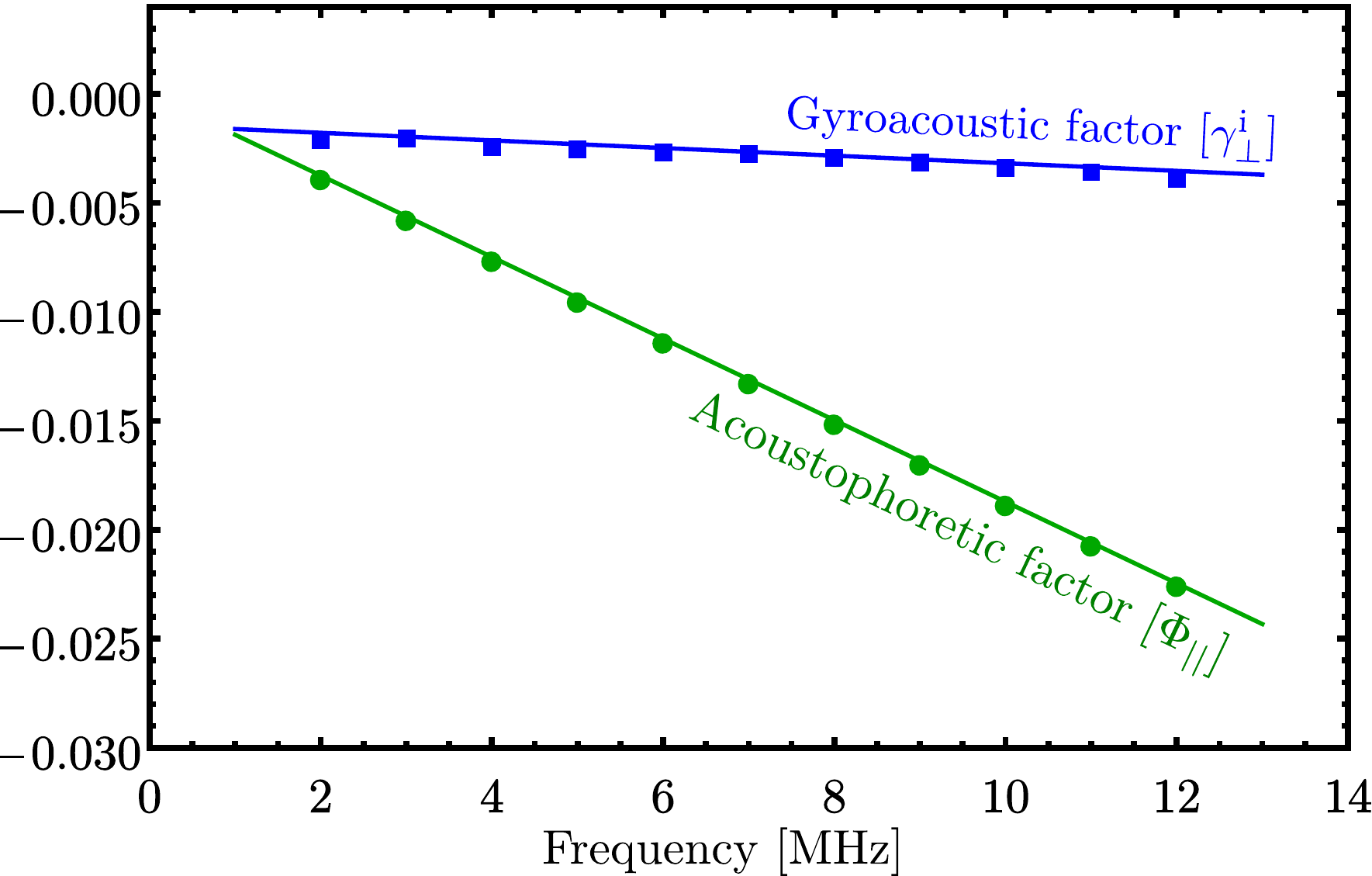}
    \caption{
    \label{fig:RFRTRBC}
    The gyroacoustic and acoustophoretic factors of a RBC in blood plasma versus frequency.
    The green and blue lines represent a linear interpolation of data.}
\end{figure}

In Fig.~\ref{fig:RFRTRBC}, we show the computed gyroacoustic and acoustophoretic factors for the RBC.
The obtained data is interpolated with polynomial functions of linear frequency $f$.
Let us first examine the gyroacoustic factor.
Referring to~\eqref{RTSW2}, we find 
\begin{equation}
{ 
    \gamma_\perp^\text{i} = - \num{0.001433} - (\SI{0.000174}{\per\mega\hertz})\, f,
    }
\end{equation}
in which the RBC volume\cite{Evans1972}  $V=\SI{94}{\micro\meter\cubed}$ was used, and the frequency should be specified in megahertz.
The normalized root mean square error between the fitting polynomial and data is 
$\text{nmrs} = 5\,\%$ (normalized to the data range).
For a typical value of the energy density in acoustofluidics, say $\SI{10}{\joule\per\meter\cubed}$, the radiation torque peak is $\SI{1.62}{\pico\newton\micro\meter}$.
As $\gamma_\perp^\text{i}$ is negative, we see the radiation torque aligns the RBC (axis of symmetry) with the axial direction ($\beta=0$). 
This prediction was experimentally observed in a half-wavelength acoustofluidic device.\cite{Jakobsson2014}
Note also that when the second orientation angle is $\alpha=0$, the radiation torque direction follows
 $\vec{\tau}^\text{rad}\sim -\vec{e}_y$.
As the l-frame is a right-handed coordinate system, the RBC rotates around its transverse diameter ($y$ axis) in the clockwise direction.
On the other hand, when
$\alpha=\pi/2$ we have 
$\vec{\tau}^\text{rad} \sim \vec{e}_x$, the RBC
 rotates around its transverse diameter,
 but this time in counterclockwise direction.
%
% It worth saying that axisymmetric particles
% seem to have a tendency to be aligned  with their largest cross-section area perpendicular to the wave axis.
% % 
% This is the case of prolate spheroidal particles\cite{Leao-Neto2020} and RBCs.

Having determined the RBC orientation,
we know the radiation force by the standing wave is described by Eq.~\eqref{FSW}.
Using a linear interpolator, we obtain the acoustophoretic factor as
\begin{equation}
{ 
    \Phi_{||} = -(\SI{0.001871}{\per\mega\hertz})\, f.
}
\end{equation}
where the transverse cross-section area $A_\perp = \pi R_0^2$ was used, and
the frequency should be given in megahertz.
The interpolation error reads
$\text{nrms}=0.1\,\%$. 
The peak radiation force for $E_0=\SI{10}{\joule\per\meter\cubed}$ and $f=\SI{2}{\mega\hertz}$ is
$\SI{1.79}{\pico\newton}$.
Moreover, as $\Phi_{||}<0$, we conclude the RBC is trapped in a pressure node, $k z_0=(2n+1)\pi/2$, with $n\in \mathbb{Z}$,
which agrees with previous experimental observations.~\cite{Jakobsson2014}

Finally, our results can be compared with those obtained by the Born approximation\cite{Jerome2019a} for the RBC at $\SI{2}{\mega\hertz}$, under the same conditions as specified in Table~\ref{tab:parameters}.
For the acoustic radiation force, the relative deviation between the results is a mere $0.4\,\%$.
In the case of the radiation torque, we cannot, unfortunately, draw a comparison because the reference only provides the torque in a pressure antinode, e.g., $kd = 0$ in the article's notation.
At this position, our method predicts a zero radiation torque.
For the sake of curiosity, we calculate the peak torque from Ref.~\onlinecite{Jerome2019a} as 
$\sim 10^{-5} \tau_0$.
This value is two orders of magnitude smaller than our outcome for the RBC in a pressure node, 
{
 
which can be considered close to our zero-torque prediction at a pressure antinode.
}

\section{Summary and conclusions}
This article introduces a semi-analytical method to compute the mean-acoustic fields (radiation force and torque) exerted on a subwavelength axisymmetric particle in a Newtonian fluid.
The analytical part draws on the partial wave expansion of the scattering pressure up to the dipole approximation.
The mean-acoustic fields are derived as a function of the scattering coefficients--see Eqs.~\eqref{Frad3} and \eqref{radtorque_wave}.
The monopole and dipole scattering coefficients are computed as the projection of the scattered pressure onto the angular part of the corresponding multipole mode--see Eq.~\eqref{s-numerical}.
These coefficients are then numerically obtained using the finite element (FE) method to solve the plane wave scattering by the particle.
{ 
    We should remember that the scattering coefficients of spherical particles depend intrinsically on the mechanical properties of the particle and surrounding fluid, regardless the incoming ultrasonic wave.
    This can be seen, for instance, in the case of solid elastic particles.\cite{Baresch2013a}
    In our approach, the scattering coefficient of axisymmetric particles are also independent of the incident wave--see Eq.~\eqref{psc}. 
    So, after obtaining these coefficients from the plane wave problem,
    one can compute the mean-acoustic fields on the particle caused by any structured wave generated in acoustofluidic devices, using Eqs.~\eqref{Frad4} and \eqref{radtorque_wave}. 
}

{ We performed numerical tests to obtain the scattering coefficients of a small rigid sphere in water.
The results are in excellent agreement with the well-known exact scattering solution.}
The semi-analytical method is showcased for a RBC subjected to a standing plane wave in blood plasma.
Our predictions for the particle orientation and entrapment location agree with previous experimental results.~\cite{Jakobsson2014}

Lastly, our work is a concrete step toward computing the mean-acoustic fields on nonspherical particles in more realistic acoustofluidics settings. 
The method is valid for particles made of fluid, elastic, viscoelastic, and structured material.
Besides, thermoviscous properties of the surrounding fluid can be incorporated into the numerical model.
Not to mention that other numerical techniques such as the boundary element method, T-matrix, and finite differences can be used here.
In conclusion,
our theory may serve as the foundation for new investigations of RBCs and elongated cell dynamics in acoustofluidic devices and also the development of micro/nanorobots propelled by ultrasound.

\begin{acknowledgments}
We thank the Brazilian National Council for Scientific and Technological Development--CNPq, grant number 
308357/2019-1. 
\end{acknowledgments}

\appendix

\section{Particle versus wave frame of reference}
\label{app:rotation}

The inverse rotational tensor, which transforms a vector from the l-frame to the p-frame, is given by    
\begin{align}
    \nonumber
    \mathbf{R}^{-1} &=
    \cos\alpha \cos\beta\, \vec{e}_{x_\text{p}}\vec{e}_x
        +\sin \alpha\cos\beta \, \vec{e}_{y_\text{p}}\vec{e}_x
        -
        \sin\beta\, \vec{e}_{z_\text{p}}\vec{e}_x
        \\
        \nonumber
        &-\sin\alpha\, \vec{e}_{x_\text{p}}\vec{e}_y 
        +
        \cos \alpha \, \vec{e}_{y_\text{p}}\vec{e}_y
        -\cos\alpha \sin\beta\, \vec{e}_{x_\text{p}}\vec{e}_z 
        \\
        &+
        \sin \alpha \sin\beta\, \vec{e}_{y_\text{p}}\vec{e}_y
        +
        \cos\beta\, \vec{e}_{z_\text{p}}\vec{e}_z.
\end{align}
The gradient operator is transformed as
\begin{subequations}
    \begin{align}
        \label{nabla}
        \nabla  &= \mathbf{R}\cdot\nabla_\text{p}|_{\vec{r}_\text{p}=\vec{r}}, \\
        \nabla_\text{p}  &= \mathbf{R}^{-1}\cdot\nabla|_{\vec{r}=\vec{r}_\text{p}}.
        \label{WPnabla}
    \end{align}
\end{subequations}

\section{Transverse dipole symmetry}
\label{app:symmetry}
We want to demonstrate that the transverse dipole mode of axisymmetric particles are degenerated, e.g., $s_{11}=s_{1,-1}$.
Assume the plane wave travels along the $x_\text{p}$ axis, then
the wavevector is $\vec{k}=k \vec{e}_{x_\text{p}}$ and $\beta=\pi/2$.
We note the corresponding beam-shape coefficients are\cite{Colton1998}
$    a_{1,-1} = -a_{11}= \ii \sqrt{6 \pi}.
$
From Eq.~\eqref{pnm}, we see that the normalized pressure  of the dipole mode satisfies
$    p_{1,-1} = - p_{11}.
$
Thus, using the spherical harmonic symmetry property
$Y_n^{-m} = (-1)^m Y_n^{m*}$ and referring to Eq.~\eqref{snm}, we have
\begin{equation}
    s_{1,-1} = 
    \int_0^{2\pi}\dd\varphi_\text{p} \int_0^\pi
    \dd \theta_\text{p}
    \, \sin \theta_\text{p} \,
    {p}_{11}(\theta_\text{p},\varphi_\text{p})
    Y_1^{1}(\theta_\text{p},\varphi_\text{p}).
     \label{s1m1}
\end{equation}
Inasmuch as the particle rotation symmetry around $z_\text{p}$ axis, the scattered pressure has to be an even function of the azimuthal angle $\varphi_\text{p}$.
Hence ${p}_{11}(\theta_\text{p},-\varphi_\text{p})={p}_{11}(\theta_\text{p},\varphi_\text{p})$, otherwise the scattered pressure will not be symmetric with respect to the $x_\text{p}z_\text{p}$ plane as it should be.
Equation~\eqref{s1m1} can be re-written as
\begin{align}
    \nonumber
    s_{1,-1} &= 
    \int_0^{2\pi}\dd\varphi_\text{p} \int_0^\pi
    \dd \theta_\text{p}
    \, \sin \theta_\text{p} \, {p}_{11}(\theta_\text{p},\varphi_\text{p})
    Y_1^{1*}(\theta_\text{p},\varphi_\text{p})
   \\
     &=\braket{1,1|{p}_{11}} =s_{11}.
\end{align}
Here  $ Y_1^{1}(\theta_\text{p},-\varphi_\text{p})= Y_1^{1*}(\theta_\text{p},\varphi_\text{p})$ was used.
We conclude the transverse dipole modes are indeed degenerated for axisymmetric particles.

%\bibliographystyle{jasanum2}
%\bibliography{Paper.bib}

\end{document}